\documentclass[10pt, conference]{IEEEtran}

\PassOptionsToPackage{hyphens}{url}
\usepackage[colorlinks = true,
  linkcolor = blue,
  citecolor = blue,
  urlcolor = blue]{hyperref}

\usepackage{url}
\usepackage{amsmath}
\usepackage{amssymb}
\usepackage{csquotes}
\usepackage{multirow}
\usepackage{wrapfig}
\usepackage{amsmath}
\usepackage{graphicx}
\usepackage{tabularx}
\usepackage{booktabs}
\usepackage{balance}
\usepackage{threeparttable}

\usepackage{fancyhdr} 
\fancypagestyle{firststyle}
{
\fancyhf{}
\fancyfoot[C]{\scriptsize{Proceedings of the Seventh International Conference on Intelligent Data Science Technologies and Applications (IDSTA 2026), Bucharest, IEEE, 2026, pp.~11--17. \\ This version is the author's copy. The publisher's definite version is available online via \url{https://doi.org/10.1109/IDSTA70607.2026.11771624}.}}}

\begin{document}

\title{Chinese Whispers About Vulnerabilities}

\author{
\IEEEauthorblockN{Jukka Ruohonen}
\IEEEauthorblockA{University of Southern Denmark \\
Email: juk@mmmi.sdu.dk}
}

\maketitle

\begin{abstract}
The dissemination of information about known software vulnerabilities occurs
online in various ways. Everything from email and development systems to social
media are used. Given this variety and a dataset of about 382 thousand online
references for about 55 thousand vulnerabilities, the paper demonstrates that
the referenced online information sources tend to be mostly dissimilar
statistically. The mean cosine similarity between them is only
0.22. Interestingly, the amounts of online references neither increase nor
decrease similarity between the online information sources. In contrast,
vulnerabilities primarily associated with GitHub and programming language
ecosystems tend to have more dissimilar online references than other
vulnerabilities. Also vulnerability reports, whether in bug tracking systems or
bug bounty platforms, tend to increase dissimilarity. With these empirical
results, the paper contributes to the efforts to narrow an enduring knowledge
gap related to the online tracking and associated traceability of software
vulnerabilities.
\end{abstract}


\section{Introduction}

\thispagestyle{firststyle} 

In recent years, there has been an increasing interest for better open source
intelligence (OSINT) in relation to cyber threat intelligence (CTI). A good
example would be data mining of vulnerability and threat information posted on
social median, including X/Twitter in particular~\cite{Syed25,
  Tundis20}. However, data quality remains a severe issue. The problem is
acknowledged by many papers operating in the OSINT-based CTI domain that have
relied on a definition emphasizing data that is relevant, actionable, and
valuable~\cite{Tundis20, Ruohonen26ICEDEG, Sun23}. As everyone knows, social
media platforms---and perhaps the whole Internet in general, are full of
misinformation, disinformation, and generally low-quality content, including
increasing amounts of misleading content generated by artificial intelligence
(AI)~tools.

When it comes to software vulnerabilities, it is also well-known that numerous
distinct software development systems and platforms are used. Despite the
emergence of GitHub and other development platforms in the late 2000s and the
arrival of what have been called social coding and social
computing~\cite{Dabbish12}, information about software vulnerabilities still
appear also on numerous other online venues, including, but not limited to,
security mailing lists~\cite{Iannone24, Ruohonen18IST} and bug bounty
platforms~\cite{Ayala25}. The diversity of the venues makes OSINT-based CTI
challenging. Another challenge is about the previous point; low-quality content
and even spam are well-known challenges also for bug bounty
and other \text{platforms~\cite{Ayala25, Bressers25, RedMonk26}}.

These two challenges motivate the first research question (RQ$_1$): \textit{how
  similar or dissimilar different online information sources about software
  vulnerabilities are?} The second research question (RQ$_2$) follows:
\textit{what can statistically explain the similarities or dissimilarities and
  how well?} Regarding RQ$_2$, the wording in the paper's title provides the
underlying \textit{hypothesis}; the more there are online sources discussing a
software vulnerability, the more dissimilar the information in overall. In other
words, Chinese whispers refer to a popular children's game in which a message is
whispered from a person to another and the message is then compared with an
assumption that it has changed while traveling between the whispering
persons~\cite{Lyytimaki14}. That is to say, the game is a metaphor for
miscommunication~\cite{Jolak20}. The children's whispering game, which is also
known as a transmission chain~\cite{Mesoudi07}, has been used also to theorize
information and its flow. For instance, it has been assumed that messages change
over time and errors accumulate, and complexity of the topic discussed tends to
increase the changes and error accumulation~\cite{Lyytimaki14}. This example
provides a further motivation for~RQ$_2$ because complexity certainly
characterizes also software vulnerability~information.

The dataset examined is based on vulnerability information extracted from the
distributed Open Source Vulnerabilities (OSV) database. In particular, the focus
is on the online references archived to the OSV database. As described in the
opening Section~\ref{sec: related work}, these can be seen as traceability links
to additional, unstructured information about the vulnerabilities archived to
the database. Then, data and methods are elaborated in Section~\ref{sec: data}
and Section~\ref{sec: methods}, respectively. Results follow in
Section~\ref{sec: results}. The last two Sections~\ref{sec: conclusion} and
\ref{sec: limitations} present the conclusions reached, a summary of
limitations, and some points about further relevant research potential along the
way.

\section{Related Work}\label{sec: related work}

``\textit{How can we uniquely identify and track a vulnerability across the
  entire history of an ecosystem?}'' \cite[p.~144:10]{Bohme25}. This question
was asked in a recent survey paper presenting a road map for addressing grand
challenges in software security research for 2030 and beyond. Although the
question is valid and as important as ever, it could be improved by using a
plural form for the noun \textit{ecosystem}. The reason is that there are
multiple software ecosystems. Therefore, it is not surprising that there has
been an increasing interest in \textit{cross-database}~\cite{Zhu25, Boles25},
\textit{cross-ecosystem}~\cite{Iannone24, Ruohonen25ICTSSb, Xu22}, and generally
\textit{cross-source}~\cite{Le22} vulnerability research. Of this almost
entirely empirical research, duplicate detection and similarity
analysis~\cite{Zhu25, Dunlap24}, online discussions~\cite{Syed25, Iannone24,
  Ruohonen18IST}, and information propagation across
ecosystems~\cite{Ruohonen25ICTSSb} are worth singling out as the closest
reference points of related work.

The verb \textit{track} in the earlier quote can be used to connect the paper
further to the research area of software \textit{traceability}. Although no
universally accepted definitions exist, software traceability has usually been
understood in graph theoretical terms; \textit{tracing} is possible between a
software artifact and another software artifact whenever there exists either a
directed edge or an undirected edge between them~\cite{Maletic05}. A specific
term \textit{traceability link} is sometimes used in this regard.

The artifacts traced vary from a paper to another and an application domain to
another. A classical example would be requirements and their traceability to
software design choices, models, and architectures~\cite{Tang11}. Traceability
links between security advisories and vulnerability fixing commits would be
another example~\cite{Dunlap24}. Given the paper's open source software~(OSS)
context, a further example would be traceability links between original and
forked OSS projects~\cite{Businge23}. All three examples are about traceability
between different artifact types; a requirement differs from a design choice, a
natural language security advisory is not a code commit, and a fork is different
from the original. Therefore, a distinction between within-artifact (or
intra-artifact) and between-artifact (or cross-artifact) traceability has
sometimes been used~~\cite{Jaber13}. Perhaps little confusingly, the terms
horizontal traceability and vertical traceability, respectively, have also been
used for conveying the same analytical meaning; the latter is about linking
artifacts at differing abstraction levels~~\cite{Gotel11a}. As the earlier
remarks about cross-database, cross-ecosystem, and cross-source research
pinpoint, the paper operates in the (vertical) \textit{cross-artifact
  traceability} branch: the artifacts observed are vulnerabilities reported and
online sources discussing the vulnerabilities, and the traceability links
between them are taken as~undirected.

Because the vulnerabilities observed are all about already released software,
the paper is also about \textit{post-release} traceability
research~\cite{Ruohonen25ICTSSb}. Regarding the traceability links, these refer
to conventional hyperlinks used in the Web. Following hyperlinks archived to
vulnerability databases has also been a way to assemble vulnerability datasets
for research purposes~\cite{Wang24}. Also hyperlinks from other sources, such as
security mailing lists~\cite{Iannone24, Ruohonen18IST}, have been traced.
However, and unlike with web crawling, the \textit{depth} of the tracing is kept
minimal; the traceability links are between the OSV database entries and online
information sources about the entries stored to the database. In other words,
tracing is not done from the online sources to further online sources via
tracing hyperlinks.

\section{Data}\label{sec: data}

The dataset, which is available online for those
interested~\cite{Ruohonen26DSCW}, was collected on the first of October 2025
from the OSV database launched and maintained by Google and
associates.\footnote{~\url{https://osv.dev/}} Although fairly new, the OSV
database has already been used in recent empirical research~\text{\cite{Zhu25,
    Boles25, Ruohonen25ICTSSb, Dunlap24, Ruohonen25ARES}}. It has also been
praised by practitioners~\cite{Bressers26}. The database's distributed nature is
perhaps its primary selling point; it curates vulnerability information from
various OSS sources, including all the major Linux distributions and programming
language ecosystems. While recognizing the OSV's terminological oddity regarding
the noun ecosystem~\cite{Ruohonen25ICTSSb}, all entries were retrieved
separately from $38$ ``ecosystems''; only something called \texttt{[EMPTY]} was
excluded.\footnote{~\url{https://osv-vulnerabilities.storage.googleapis.com/ecosystems.txt}}
With these thirty-eight ``ecosystems'', the raw data includes over $422$
thousand unique vulnerability entries delivered in the JavaScript Object
Notation (JSON) format. The raw data reduced to $n = 55,224$ entries after
pre-processing.

Four pre-processing restrictions were imposed upon the vulnerability entries
qualified into the sample analyzed: (1)~only vulnerabilities with Common
Vulnerabilities and Exposures (CVE) identifiers were included in the sample;
(2)~the CVE-referenced vulnerability entries were required to have Common
Vulnerability Scoring System (CVSS) severity entries directly available from the
OSV database; (3)~only vulnerability entries with at least two references to
external sources were included; and (4) only those vulnerabilities were included
for which at least two external online sources were successfully retrieved. In
practice, the CVE identifiers were retrieved from the \texttt{id} field in the
OSV's JSON schema. For six ``ecosystems'', including Debian and Ubuntu,
``ecosystem-specific'' prefixes, such as \texttt{DEBIAN-} and \texttt{UBUNTU-},
were removed from the CVE identifiers retrieved from the \texttt{id} field. If
an invalid CVE identifier was still present, as determined by checking whether a
given string started with a \texttt{CVE-} substring, the \texttt{aliases} field
was used, provided that only a single alias was present and it started with the \texttt{CVE-} substring. All CVE identifiers were further checked to have
years between 1990 and 2025.

The first restriction about CVEs, which has been commonly imposed in existing
research~\cite{Syed25, Iannone24, Ruohonen25ICTSSb, Wang24}, can be justified on
the grounds that the references observed are about publicly recognized and more
or less well-known vulnerabilities. The second restriction further reinforces
this assertion. Because poorly known, poorly reported and coordinated, or even
obscure vulnerability entries are excluded, the references can be expected to
exhibit similarities already because the amount of vague and unverified sources,
such as rumors and speculations on social media, can be assumed to be
limited. The second restriction further reinforces the assumption. Regarding the
third restriction, it should be clarified that also entries for which less than
two entries were successfully retrieved were excluded. Although rigorous
record-keeping was not done, various errors occurred, ranging from dead
hyperlinks to tracking systems and platforms only available with JavaScript or
those imposing restrictions upon crawling. Particularly the so-called
\textit{link rot} observed would warrant further research because it seems
traceability may worsen over time, as hinted by existing
studies~\cite{Ruohonen26ICEDEG, Hata19}. AI may make the situation
even~worse~\cite{Bednar26}.

Finally, regarding the content downloaded, all Hypertext Markup Language
elements were removed but no other processing was done due to the numerous
distinct online sources referenced in the OSV's traceability links. As some
recurring boilerplate text occurs in many of the online sources referenced, such
as those pointing to GitHub or open bug trackers, the text similarity metrics
soon described in Subsection~\ref{subsec: metrics} can be assumed to have lower
bounds; their values often never reach the metrics' theoretical minimums in the
present context

\section{Methods}\label{sec: methods}

In what follows, the metrics and their operationalizations are first
elaborated. Subsequently, the statistical regression methodology for them is
outlined.

\subsection{Metrics}\label{subsec: metrics}\label{subsec: metrics}

The empirical analysis operates with the following four dependent metrics:

\begin{enumerate}
\itemsep 4pt
\item{$\textmd{COSSIM}^m_i$ and $\textmd{COSSIM}^d_i$ : the mean and median
  cosine similarities between two or more online references for the $i$:th
  CVE-referenced vulnerability.}
\item{$\textmd{JACSIM}^m_i$ and $\textmd{JACSIM}^d_i$: the mean and median
  Jaccard similarities between two or more online references for the $i$:th
  CVE-referenced vulnerability.}
\end{enumerate}

Regarding these text similarity metrics, (i)~these were chosen because they are
well-known, widely used, simple, easy to implement, and easy to interpret. These
characteristics also outweighed a temptation to use large language models. The
temptation was further reduced by recent research indicating that large language
models have substantial problems in the CTI context~\cite{Chen26, Deason25,
  Meng25}. These results align with anecdotal experiences of some OSS
developers~\cite{Bressers25, Claburn26a}. A further justification is the nature
of the data; (ii)~many of the online references contain both natural language
and code snippets. While there are many metrics and algorithms to measure the
semantic similarity of natural language~\cite{Othman26}, and recent work has
extended them toward source code as well~\cite{Ren26}, it remains unclear how
well they perform with heterogenous and noisy datasets. The metrics and their
simplicity can also be justified on the grounds that (iii)~no explicitly
comparable work exists regarding RQ$_1$~and~RQ$_2$. That is, there is little
reason to prefer complex metrics without benchmarks or empirical referenced
~points. As for computation, tokenization was done according to white space, and
tokens with lengths less than three characters or more than twenty characters
were excluded. Term frequencies were used for computing the cosine similarities.

Eleven independent metrics are used to model the six dependent metrics. These
metrics can be described as follows:
\begin{enumerate}
\itemsep 4pt
\item{$\textmd{REFCOUNT}_i$: the number of successfully retrieved online
  references for the $i$:th vulnerability. Given what was noted earlier in the
  preceding Section~\ref{sec: data}, $\textmd{REFCOUNT}_i \geq 2$ holds for
  all $i = 1, \ldots, n$.}
\item{$\textmd{CVECOUNT}_i$: the number of all unique CVE identifiers occurring
  in the references for the $i$:th CVE-referenced vulnerability. A regular
  expression used in existing research~\cite{Ruohonen18IST} was used to count
  the unique CVE identifiers}
\item{$\textmd{CVESHARE}_i$: a percentage share of the occurrence of the $i$:th
  CVE identifier in all (including non-unique) CVE identifiers present in the
  references for the $i$:th CVE. For instance: if there would be five references
  for an imaginary CVE-9999-1234 but only two of them would actually contain the
  character string CVE-9999-1234, the metric would take a value $2 / 5 \times
  100 = 40$.}
\item{$\textmd{SEVERITY}_i$: the severity (CVSS~v.~3.1) of the given
  CVE-referenced vulnerability measured with a base score ranging from zero to
  ten (the higher, the more severe) and as directly reported in the OSV
  database.}
\item{$\textmd{GITHUB}_i$: a dummy variable taking a value zero unless the
  $i$:th vulnerability was categorized in the OSV database as primarily
  belonging to GitHub. While acknowledging the uncertainties regarding the OSV's
  terminology and categorizations~\cite{Ruohonen25ICTSSb}, a value one is
  recorded in case the $i$:th vulnerability was reported in the OSV's
  \texttt{GIT} or \texttt{GitHub Actions} categories.}
\item{$\textmd{PROGECO}_i$: a dummy variable taking a value zero unless the
  $i$:th vulnerability was categorized in the OSV as primarily belonging to any
  well-known programming language ecosystem in which case the metric takes a
  value one. The ecosystems are CRAN, Go, Maven, npm, Packagist, PyPI, and
  RubyGems. With the exception of Packagist for PHP, the same ecosystems have
  been examined also previously~\cite{Ruohonen25ARES}. The operationalization is
  analogous to what was described for~$\textmd{GITHUB}_i$.}
\item{$\textmd{ADVISORY}_i$: a dummy variable taking a value zero unless
  \textit{any} of the references for the $i$:th vulnerability pointed to a
  security advisory in which case the metric takes a value one. The
  operationalization is based on the OSV's \texttt{type} field provided for
  references; in this case the field's value is the \texttt{ADVISORY} character
  string.}
\item{$\textmd{BUGFIX}_i$: a dummy variable constructed analogously to
  $\textmd{ADVISORY}_i$ but for the \texttt{FIX} strings indicating a bug fix,
  as manifested by a trace in a bug tracking system, version control system, or
  other tracking system.}
\item{$\textmd{EVIDENCE}_i$: a dummy variable constructed analogously to
  $\textmd{ADVISORY}_i$ but for the \texttt{EVIDENCE} character strings
  indicating some evidence confirming the validity of the $i$:th
  vulnerability. Based on a brief qualitative outlook, these reference cases
  include proof-of-concepts~\cite{Ruohonen20CHB}, including test cases and other
  reproducible demonstrations~\cite{Ruohonen25SANER}, confirmations by OSS
  developers, detailed blog posts, and related information.}

\item{$\textmd{REPORT}_i$: a dummy variable constructed analogously to
  $\textmd{ADVISORY}_i$ but for the \texttt{REPORT} character strings indicating
  one or more references pointing to an initial report, or initial reports,
  about the $i$:th CVE-referenced vulnerability. Again based on a brief
  qualitative look, these reporting cases range from reports in bug tracking
  systems and bug bounty platforms to public posts on mailing lists, among other
  related sources.}
\item{$\textmd{YEAR}_i$: the \textit{approximate} year when a CVE identifier was
  assigned for the $i$:th vulnerability. For keeping things simple, the years
  were extracted from the CVE identifiers themselves. Given different events and
  states through which vulnerabilities can be seen to live
  through~\cite{Ruohonen20CHB, Syed25}, $\textmd{YEAR}_i$ can be interpreted to
  proxy what might be called an ``archival state''. It should be also stressed
  that the dates of the references observed may or may not equal this archival
  year. In other words, they may also be older or newer.}
\end{enumerate}

Of these metrics, particularly interest is on $\textmd{REFCOUNT}_i$.

\subsection{Regression Analysis}\label{subsec: regression analysis}

The four dependent similarity metrics are doubly truncated data: because all
their values are on the unit interval, $[0, 1]$, there are truncation points at
zero and one. These make conventional ordinary least squares regression
estimates biased~\cite{Breen96}. Therefore, the \texttt{crch} function for
censored regression (with zero and one as the left and right censoring points)
from the similarly named \texttt{crch} package~\cite{crch} for R is used for
estimation.\footnote{~Given the subtle but still important theoretical
differences between censoring and truncation~\cite[pp.~3--4]{Breen96}, it could
be argued that truncated regression would be the optimal estimation
strategy. However, the optimizations for the corresponding \texttt{trch}
function from the \texttt{crch} package~\cite{crch} result in errors for the
dataset used. Therefore, censored regression is used for practical~reasons.} Four separate censored regressions are estimated for the four
similarity metrics. The independent metrics enter into these regressions in the
order listed in the preceding Subsection~\ref{subsec: metrics}. Regarding
$\textmd{YEAR}_i$, there are as many as $20$ dummy variables capturing the years
from 2005 to 2024; the year 2025 acts as the reference.

Given the large sample size, inference with statistical significance does not
make much sense---indeed, regardless of a dependent similarity metric, the
coefficients for almost all independent metrics are statistically significant at
a 99\% level. Therefore, the regression coefficients themselves provide more
sensible interpretation material. When considering the scale of the dependent
metrics, a coefficient with an absolute value exceeding $0.05$ can be seen to
provide a ``visible effect'', that is, an effect likely distinguishable from
noise. Analogously, a threshold $0.10$ is interpreted to provide a
``notable~effect''.

\section{Results}\label{sec: results}

In what follows, a few descriptive statistics of general interest are presented
first before continuing to present and elaborate the more formal regression
analysis results.

\subsection{Descriptive Statistics}

\begin{figure}[th!b]
\centering
\includegraphics[width=\linewidth, height=9.7cm]{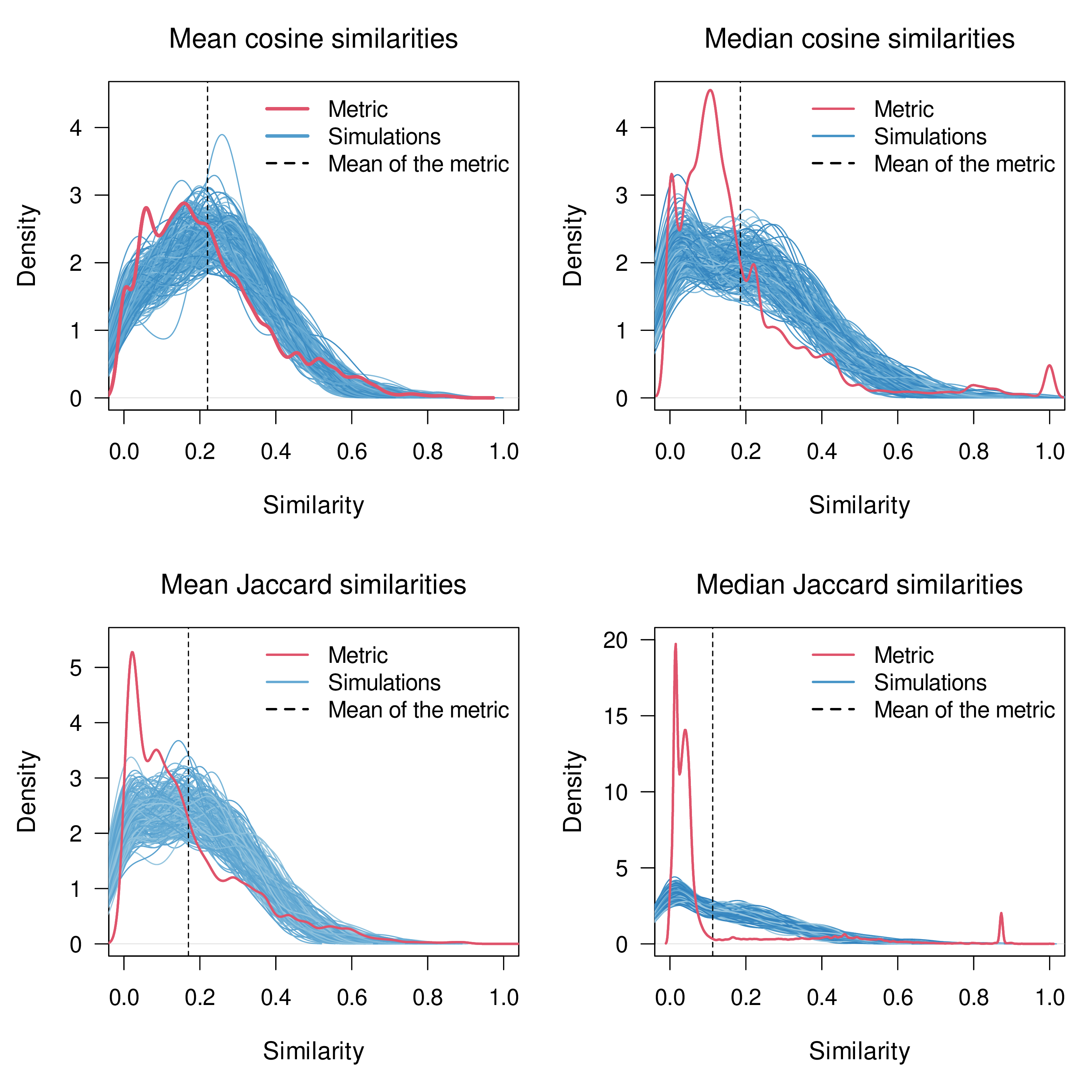}
\caption{Density Estimates of the Four Dependent Metrics with Simulations from
  the Censored Normal Distribution (means and standard deviations of the $250$
  simulated random variables with $100$ observations set to the corresponding values of
  the observed metrics)}
\label{fig: density}
\end{figure}

\begin{figure}[th!b]
\centering
\includegraphics[width=\linewidth, height=5.6cm]{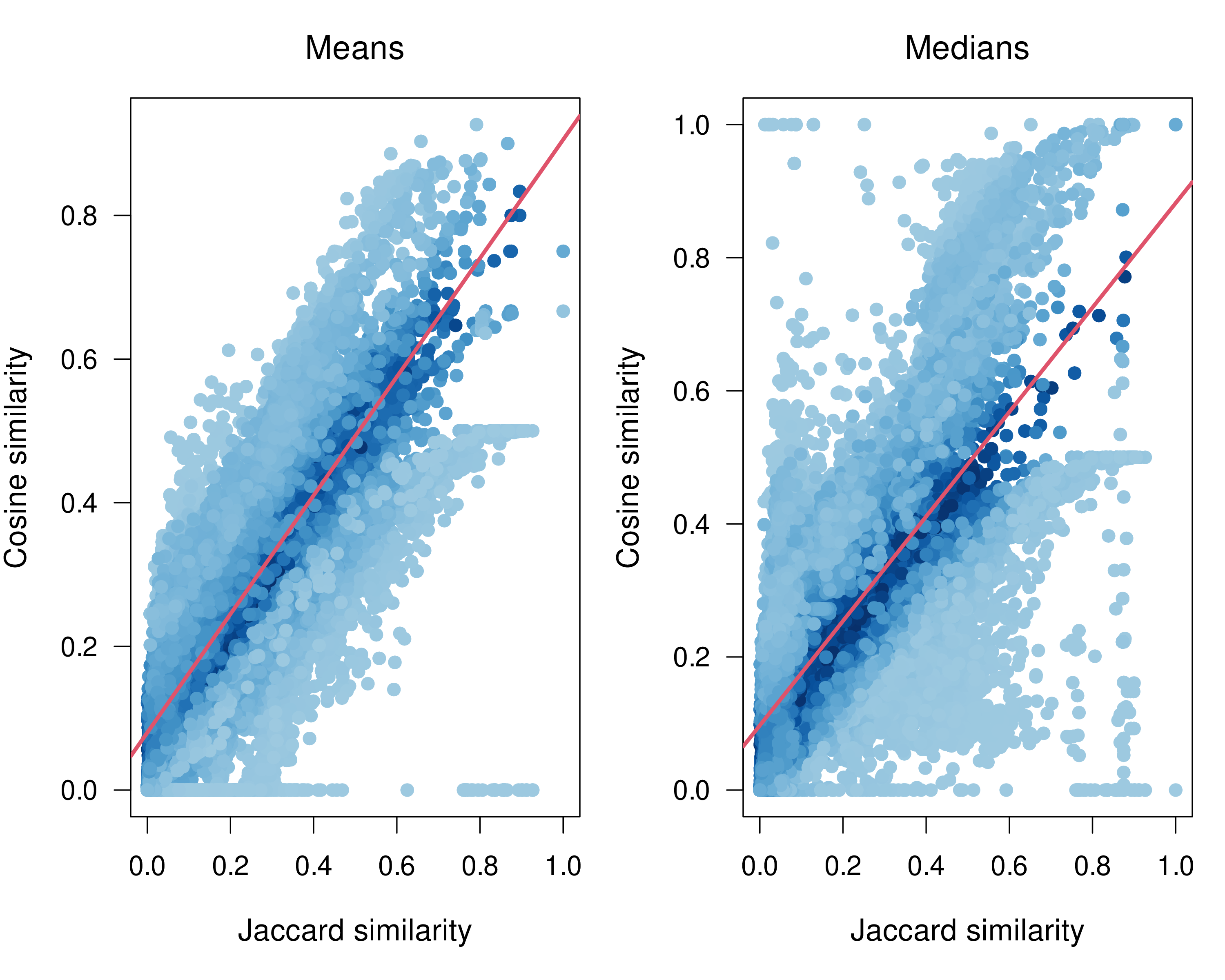}
\caption{Correlations Between the Similarity Metrics}
\label{fig: simcor}
\end{figure}

The distributions of the four dependent similarity metrics are visualized in
Fig.~\ref{fig: density}. The density estimates shown indicate that the answer to
$\textmd{RQ}_1$ is on the side of dissimilarity; the masses of all four
distributions are clearly below $0.5$. Given that as many as
$\sum^n_{i=1}\textmd{REFCOUNT}_i = 382,192$ references are present for the $n =
55,224$ vulnerabilities in the dataset, the overall dissimilarity is hardly
surprising as such. That said, the distributions have tails too; some of the
online references are quite similar. For all four metrics, the tails also range
all the way to the value $1.0$, suggesting that a few outlying duplicates are
also present. Despite the somewhat different distributional shapes, the mean
similarity metrics and the median similarity metrics correlate strongly, as can
be seen from Fig.~\ref{fig: simcor}.

Furthermore, as can be seen from the upper-left plot, the density of the
$\textmd{COSSIM}^m_i$ metric resembles quite closely the corresponding densities
of simulations from the censored normal distribution. To a lesser extent, the
densities of $\textmd{COSSIM}^d_i$ and $\textmd{JACSIM}^m_i$ too at least
resemble the shape of the simulated densities. In contrast,
$\textmd{JACSIM}^d_i$, the median Jaccard similarities shown in the bottom-right
plot, exhibits a long-tailed distribution with most of the values indicating
dissimilarity. Similar simulations with the censored logistic distribution and
the censored $t$-distribution do not resemble the empirical distribution of
$\textmd{JACSIM}^d_i$ either. Given these notes, the regression analysis soon
elaborated in Subsection~\ref{subsec: regression analysis results} is computed
with the censored normal distribution, but care should be taken when
interpreting the results for $\textmd{JACSIM}^d_i$ in particular and to a lesser
extent $\textmd{JACSIM}^m_i$ and $\textmd{COSSIM}^d_i$, respectively. In other
words, the mean cosine similarities are taken as the baseline for drawing
overall conclusions.

\subsection{Regression Analysis Results}\label{subsec: regression analysis results}

\begin{figure}[p!]
\centering
\includegraphics[width=\linewidth, height=10cm]{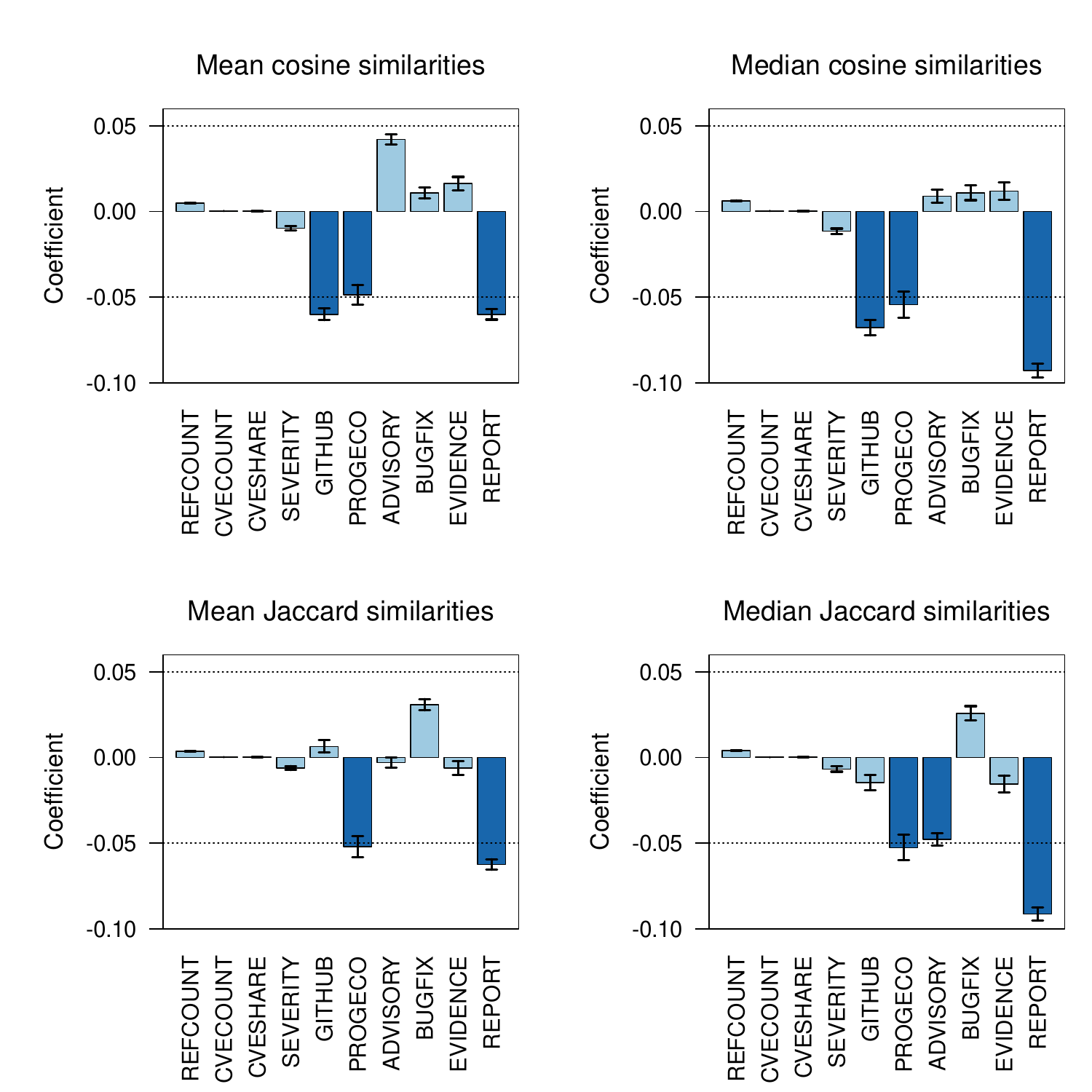}
\caption{Censored Regression Effects (95\% confidence intervals)}
\label{fig: effects}
%
\vspace{30pt}
%
\centering
\includegraphics[width=\linewidth, height=10cm]{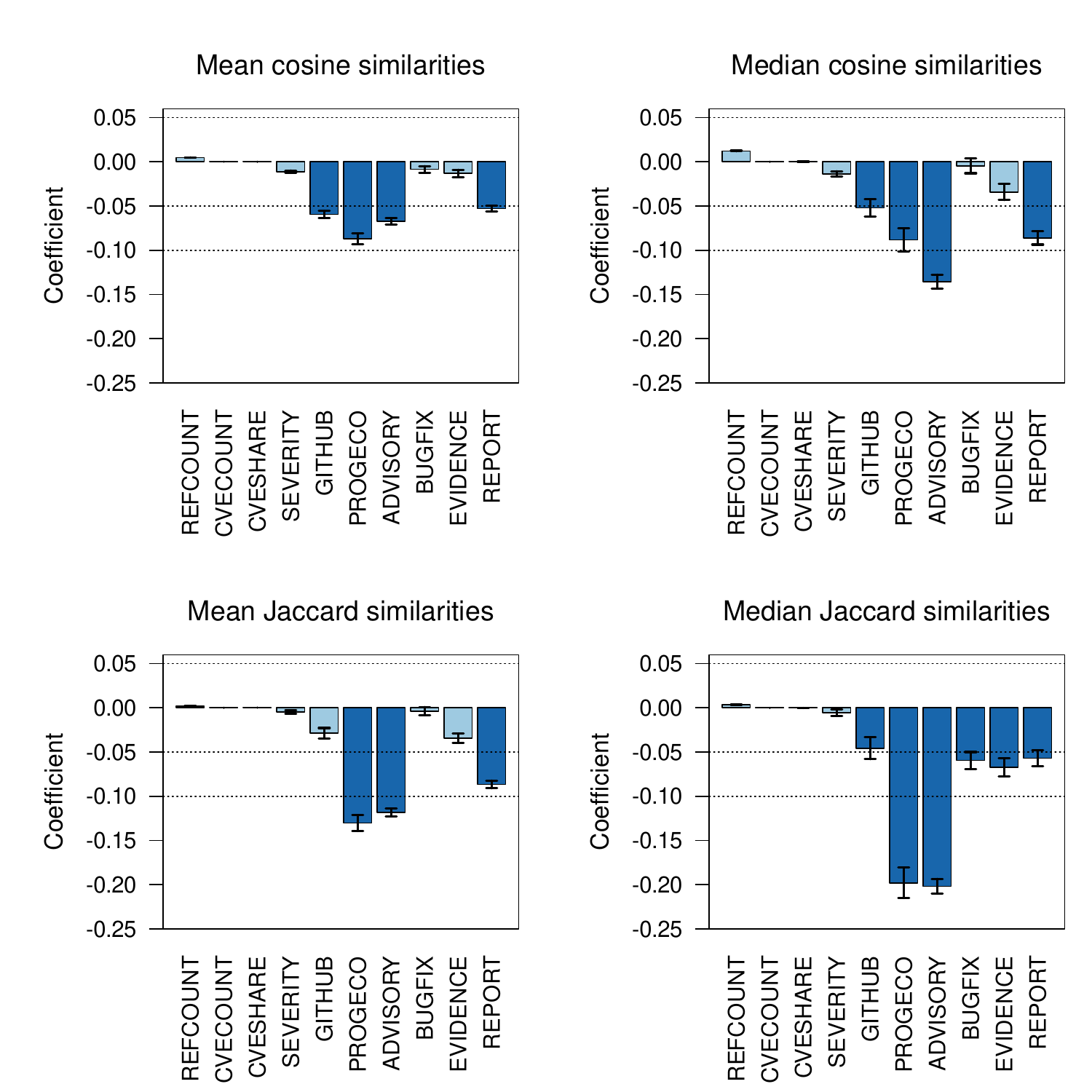}
\caption{Censored Reg.~Effects in the Similarity Subset (95\% conf.~intervals)}
\label{fig: effects simsub}
\end{figure}

The regression analysis result can be summarized by focusing on the effects, the
regression coefficient. By excluding the yearly dummy variables for brevity,
Fig.~\ref{fig: effects} visualizes the effects of the remaining independent
metrics. The first thing to note from the figure is that the prior hypothesis
does not hold; the effect of $\textmd{REFCOUNT}_i$ is small. In fact, it is
close to zero when estimation noise is taken into account. In other words, it
cannot be concluded that more online information sources about vulnerabilities
would entail more dissimilar information---or the other way around. In
contrast---and given the thresholds that were discussed in
Subsection~\ref{subsec: regression analysis}, a few of the independent metrics
do how visible effects.

\begin{figure}[th!b]
\centering
\includegraphics[width=\linewidth, height=9.5cm]{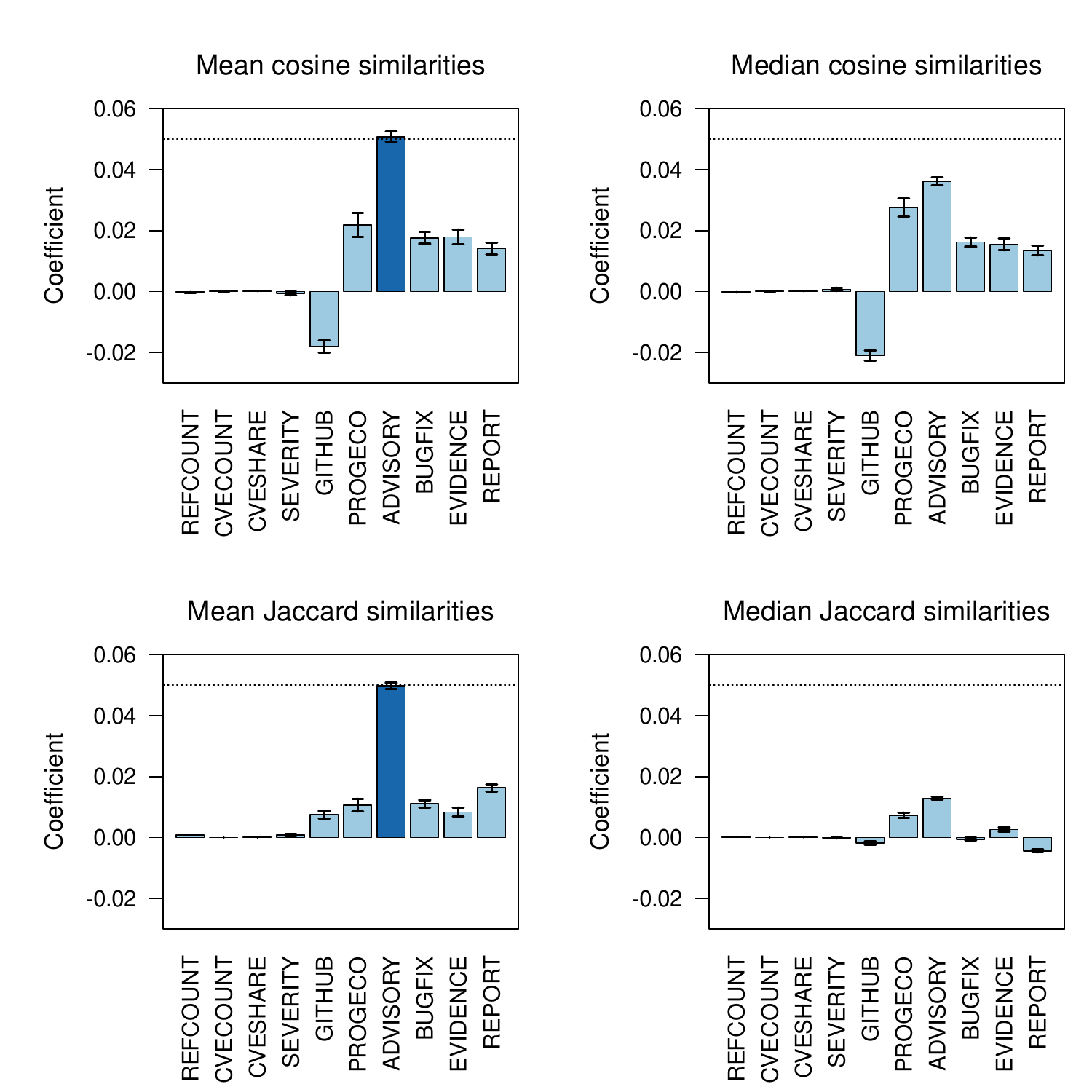}
\caption{Censored Reg.~Effects in the Dissimilarity Subset (95\% conf.~intervals)}
\label{fig: effects dissub}
\end{figure}

Even though the effects are not entirely consistent across the four
visualizations in Fig.~\ref{fig: effects}, the baseline of the mean cosine
similarities allow to group $\textmd{GITHUB}_i$, $\textmd{PROGECO}_i$, and
$\textmd{REPORT}_i$ as the independent metrics exhibiting visible effects. The
signs of the corresponding regression coefficients are all negative, indicating
a tendency to increase dissimilarity. For example, the GitHub-referenced
vulnerabilities tend to have less similar references than other vulnerabilities
in the OSV database. The observation is interesting in its own right because of
the GitHub's central role in the contemporary propagation of vulnerability
information in the OSS \text{domain~\cite{Ruohonen25ICTSSb,
    Segal26}}. Analogously, the online references for vulnerabilities associated
in the OSV database with the CRAN, Go, Maven, npm, Packagist, PyPI, and RubyGems
ecosystems tend to be slightly more dissimilar than the online references of
other vulnerabilities in the database.

To further examine the effects, additional regressions can be estimated in
``similarity and dissimilarity subsets''. To differentiate between the two
subsets, arithmetic means are used; hence, the dissimilarity subset contains
those vulnerabilities whose online references exhibit a similarity less than the
mean similarity across all vulnerabilities in the dataset. Given the dashed
vertical lines in Fig.~\ref{fig: density}, the similarity subsets thus refer to
the observations on the right of the dashed lines. With this clarification in
mind, the effects of $\textmd{GITHUB}_i$, $\textmd{PROGECO}_i$, and
$\textmd{REPORT}_i$ largely prevail in the similarity subsets. As can be further
seen from Fig.~\ref{fig: effects simsub}, also some other independent metrics
exhibit visible and even notable effects in this subset. In contrast, the
corresponding effects are absent in the dissimilarity subset shown in
Fig.~\ref{fig: effects dissub}. Within this subset, it seems
$\textmd{ADVISORY}_i$ tends to increase similarity, hinting that security
advisories may be somewhat similar, possibly due to boilerplate text used in
these. All in all, it can be concluded that statistical effects are present
mostly for those vulnerabilities whose online references are dissimilar in
overall. Although this conclusion too could be interpreted to indirectly reflect
the phenomenon of Chinese whispers, the hypothesis presented in the introduction
does not hold with any of the regressions estimated.

\section{Conclusion}\label{sec: conclusion}

The paper examined the statistical similarities and dissimilarities between
online information sources referenced in the OSV database for vulnerabilities
archived into it. The answer to $\textmd{RQ}_1$ is that the online information
sources are dissimilar in overall. For all four similarity metrics used, the
scores are below $0.23$. However, the similarity metrics exhibit also relatively
long tails; some of the online information sources are rather similar to each
other statistically. Even a few duplicates are likely present. Then, the answer
to the subsequent $\textmd{RQ}_2$ is that a few visible and even some notable
statistical effects are present for statistically explaining the values from the
similarity metrics. In particular, vulnerabilities primarily associated with
GitHub and programming language ecosystems tend to have more dissimilar online
information sources than other vulnerabilities. Also bug reports, bug bounty
reports, and other related reports tend to increase dissimilarity. However,
these effects mostly apply only to vulnerabilities whose online information
sources are generally dissimilar. Finally, the amounts of online references tend
to neither increase nor decrease similarity. Further research is required on
this regard because the observation is unexpected and feels somewhat~illogical.

\section{Limitations and Further Work}\label{sec: limitations}

Four limitations can be briefly mentioned. The first is about overall
traceability with online traceability links---or a lack thereof. That is, a
large amount of the online references archived to the OSV database could not be
successfully retrieved. This observation is familiar also from existing
studies~\cite{Dunlap24}. Link rot applies also to vulnerability databases and
CTI sources~\cite{Ruohonen26ICEDEG}, and \textit{traceability decay} is among
the grand challenges of traceability research~\cite{Gotel11b}. The second
limitation is related. Although a large dataset was examined, generalizability
remains a question mark---as almost always in this field of
research~\cite{Iannone24}. The third limitation is about the similarity metrics
used. Although the arguments raised in Subsection~\ref{subsec: metrics} are
reasonable for the present work, semantic text similarity metrics could provide
further insights and reassure the validity of the results reported. That said,
the fourth limitation is the sole reliance on text similarity metrics.

As for the last limitation, human evaluation would be required too for deducing
whether the online information sources referenced are actually dissimilar in a
sense of being valuable, providing further actionable information not otherwise
available. The last limitation is important to emphasize because so-called AI
\textit{slop} is currently flooding on GitHub and other platforms, causing
maintenance issues for OSS \text{projects~\cite{RedMonk26, Baltes26}}. By
hypothesis, slop is mostly similar.

Therefore, an evaluation with human subjects might allow to also deduce about
whether and to which extent dissimilarity is a good think for OSINT about
software vulnerabilities. If information would be seen by human subjects to be
mostly similar, it would seem that redundancy---if not outright duplication---is
present among the various cyber threat intelligence feeds. This hypothesis has
recently been raised also among practitioners~\cite{Bressers26}. If that would
indeed be the case, it could be further argued that practical curation of
vulnerability OSINT should focus on dissimilar sources and
references. Redundancy is a poor rationale for curation.

\balance
\bibliographystyle{IEEEtran}


\end{document}